\documentclass[%
 reprint,
 amsmath,amssymb,
 aps,
]{revtex4-2}

\usepackage{graphicx}
\usepackage{dcolumn}
\usepackage{bm}
\usepackage{subfigure}
\usepackage[dvipsnames]{xcolor} 
\usepackage{ulem}

\DeclareMathOperator{\erf}{erf}

\begin{document}

\preprint{APS/123-QED}

\title{Tunable effective diffusion of CO$_2$ in aqueous foam}

\author{Cécile Aprili}
\author{Gwennou Coupier}
\author{\'Elise Lorenceau}
\author{Benjamin Dollet}
 \email{benjamin.dollet@univ-grenoble-alpes.fr}
\affiliation{%
 Univ. Grenoble Alpes, CNRS, LIPhy, 38000 Grenoble, France
}%


\begin{abstract}
Aqueous foams are solid materials composed of gases and liquids, exhibiting a large gas/liquid surface area and enabling dynamic exchanges between their fluid components. The structure of binary-gas foams, whose bubbles consist of a mixture of two gases having different affinities with the liquid, thus offers real potential for the dynamic separation of these gases at low cost. In single-gas foams, the foam structure evolves under the effect of gas flow induced by Laplace pressure differences, arising from heterogeneities in bubble size. This leads to the well-documented Ostwald ripening. In addition to these capillary effects, the structure of binary-gas foams can evolve under the effect of gas flow induced by partial (or osmotic) pressure differences, arising from heterogeneities in bubble composition. We experimentally investigate the shrinking of CO$_2$-laden 2D foams exposed to air, observing a crust of tiny bubbles at the front. We derive a non-linear diffusion model for the gas in the foam and propose a description of the whole foam as an effective, homogeneous medium, the key parameter being the gas permeability ratio across the foam's soap films ($\neq 1$ for CO$_2/$air). The effective diffusivity of the gas in the foam emerges from the coupling between foam structure and gas transport across soap films. We extrapolate it for various permeability ratios and show that it can vary continuously between the diffusivity of the gas in the liquid and that of the gas in the atmosphere, enabling tunable gas retention and release by controlling the composition of the atmosphere.
\end{abstract}

\maketitle


\section*{Introduction}

Reducing the cost and energy required to set up efficient separation, sorting and filtration processes, with the aim of limiting the consumption of natural resources, is one of the key challenges facing our societies in terms of sustainable development \cite{Sholl2016}. In this context, the separation of different gases from a mixture at low energy and resource cost is particularly sensitive given the urgent need to reduce capturing carbon dioxide (CO$_2$) emissions. Membrane-based separation methods, or other non-thermal methods, can be significantly less energy consuming than heat-based separations, which are nevertheless the main ones used today \cite{Sholl2016}. Separation of the different gases in a gas mixture is generally carried out as follows \cite{Aaron2005}. First, the mixture is brought into contact with a substrate that can adsorb (then the substrate is usually solid) or absorb (then the substrate is usually liquid) the gas of interest such as CO$_2$, the other gases being released into the atmosphere. Once this separation step has been completed, the solid or liquid substrate is regenerated, i.e. the sorbed gas is released and reused or stored in the ground enabling the start of a new separation cycle \cite{Hepburn2019,Aminu2017}. An important difference between these two types of substrates, i.e. solid or liquid, concerns the question of regeneration: whereas solid substrates must be regenerated by temperature or pressure swing cycles, liquid substrates can be regenerated continuously using a flow, thus enabling continuous operation. In this case, the gas-liquid interface or specific interfacial area must be maximized to optimize transfer kinetics. This is generally achieved by using gas/liquid contactors, which are often porous solids on which the solvent flows by gravity, while the gas is pushed by a pump \cite{Tan2012}.

Although these contactors enable a large specific interfacial area, they should also be selective toward the gas of interest as well as highly permeable to the other components of the initial gas mixture to ensure high separation throughput. 
Therefore, high selectivity, large specific area and gas permeability, combined with the ability to work continuously, are the essential requirements to achieve efficient gas/liquid mass transfer for gas separation application.
The structure of aqueous foams, well documented in the literature \cite{stevenson,cantat}, seems to meet these criteria: the continuous liquid network of channels (so-called Plateau borders) and nodes around the bubbles allow continuous regeneration of liquid, while the specific interfacial area of foam columns can be several times greater than that of a packed column \cite{stevenson,Charpentier1981}. Foams also present the advantage to be self-supported structures, i.e. the interfacial areas are the structuring elements of the system. This simplicity opens the way towards low-cost design of gas separators \cite{Ashrafizadeh2024} but brings an additional difficulty: the structure of the separator will depend on the separation process itself, through a modification of the gas concentrations in the bubbles, therefore of their partial pressures across the films.

 \begin{figure*}
\centering
\includegraphics[width=1\linewidth]{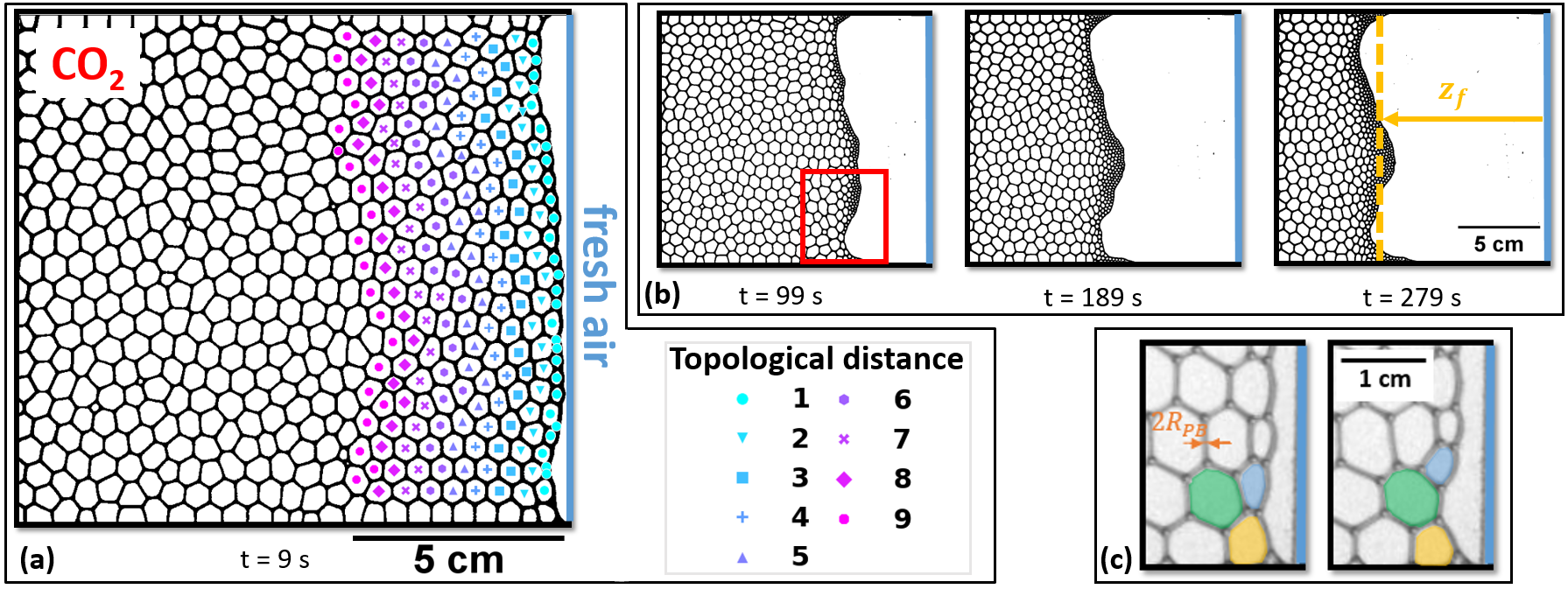}
\caption{\label{img:evolution:foam} (a) Top view snapshot at $t = 9$~s of a 2D foam initially filled with CO$_2$ confined within a Hele-Shaw cell, and exposed to air at the open end (sketched as a blue line at the right end of the snapshot) starting at an initial time $t = 0$. The two sides walls correspond to the top and bottom black lines. The bubbles pertaining to the nine first layers from the front are labeled according to their topological distance. (b) Evolution of the foam at three different times; the snapshot at $t = 279$~s illustrates the mean distance $z_f$ between the front and the open end of the cell. The red rectangle corresponds to the inset of Fig.~\ref{img:evolution_topological_distance}. (c) Example of a plastic rearrangement due to the shrinkage of the bubbles: the blue and yellow bubbles, both in contact with the front, lose their mutual contact between the two snapshots (separated by 0.6~s), and the green bubble initially located in the second layer inserts itself between them.
}
\end{figure*}

Regarding selectivity, the pioneering work of Princen and Mason showed a high permeability of a liquid film to pressure-driven gas transfer and a preference for CO$_2$ transfers with respect to air, owing to the large value of the ratio of the permeability of the CO$_2$ and air in the liquid, of order 50 \cite{Princen1965},  a value confirmed in later studies considering pressure-driven gas transfer in foams \cite{Saulnier2015, Forel2019, Farajzadeh2008}. Studies concerning osmotic-driven gas flow in foams \cite{Gandolfo1997, Webster2001, Yip2013, Bey2017}, which are scarce and exist mainly on the scale of soap films or bubbles \cite{Cook1974, Haas1975, Shim2014, Hadji2020, Dollet2023, Trinh2023, Hadji2024} suggest permeability contrasts between CO$_2$ and air that are lower than those observed by Princen and Mason \cite{Princen1965}. Importantly, Hadji \textit{et al.} \cite{Hadji2024} have shown that the final position of a single film separating two compartments both filled with two gases of different permeabilities depends on this permeability contrast, i.e. it is not the result of an equilibrium principle but of the full history of the diffusion processes. At the scale of a whole foam, these film displacements induce bubbles rearrangements, hence additional convective-like gas flux within the foam, together with a change in the structure of the foam itself. 

Separation optimization has thus to account for this coupled flow-structure problem, which we tackle in this work with a bidimensional (2D) foam allowing easy tracking of bubble size, for which we do not impose any liquid flow. The bubbles, which initially contain 100$\%$ CO$_2$, are exposed to ambient atmosphere. We confront the observations of the evolution of the structure of the foam with a continuum model at the foam scale: we derive a set of partial differential equations able to predict the evolution of the gas contents inside the foam over scales much larger than that of individual bubbles, in the spirit of other approaches where coarse-graining is applied over a large number of individual items contained in a representative volume element, like, in the context of foams, the theory of drainage \cite{Koehler2000,Hilgenfeldt2001}.

\section*{Results}
\subsection*{Experiment}
The 2D foam is a bubble monolayer confined between two glass plates separated by a thin gap $h = 2$~mm, with two side walls separated by a distance $w = 15$~cm (Fig.~\ref{img:evolution:foam}a). The foam is created in this  Hele-Shaw cell by bubbling CO$_2$ in a reservoir located upstream the cell and full of a solution of sodium dodecyl sulfate (SDS) at concentration 10~g/L. The downstream end of the cell is open to the atmosphere. Once the cell is filled with foam, the bubbling process is stopped and the foam is let to spontaneously evolve under the action of the gas exchanges with the atmosphere.

As time goes on, the bubbles closest to the foam/atmosphere boundary, which we call the front, start to shrink. This shrinkage then concerns deeper bubbles inside the foam, and the latter retracts inside the cell (Fig.~\ref{img:evolution:foam}a and b). We checked that throughout the experiment, even the smallest bubbles remained confined between the two glass plates, which ensures that the foam remains 2D.

\subsection*{Diffusion and shrinkage dynamics} 
\begin{figure}
\centering
\includegraphics[width=1\linewidth]{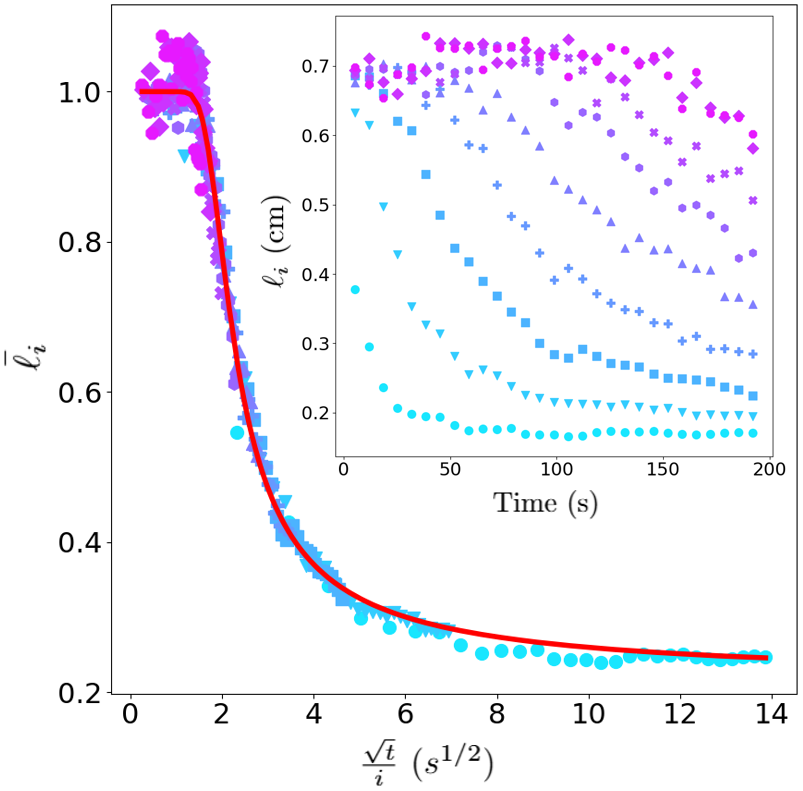}
\caption{\label{Fig:areas:fit} Inset: evolution of the average bubble size for the first nine layers in the foam, for a reference experiment. Main figure: same data, rescaled to plot $\bar{\ell}_i = \ell_i(t)/\ell_0$ as a function of $\sqrt{t}/i$. The red curve is the best fit of the rescaled data, coming from the numerical resolution of Eq. \ref{Eq:ODE_x_xi}, with the permeabilities $k_{\mathrm{air}}$ and $k_{\mathrm{CO}_2}$ as fitting parameters. Symbols refer to the same layers as in Fig. \ref{img:evolution:foam}a.}
\end{figure}

To quantify the shrinkage dynamics, we measure for each bubble, on each image, its area $\mathcal{A}$ and its topological distance $i$, defined as the minimal number of films separating the considered bubble and the atmosphere (Fig.~\ref{img:evolution:foam}a). In particular, bubbles located right at the front have a topological distance $i = 1$. The set of all bubbles sharing the same topological distance will be henceforth called the layer of index $i$. Assuming that the bubbles are in average regular hexagons (a reasonable assumption from the foam snapshots of Fig.~\ref{img:evolution:foam}), the average thickness $\ell_i$ of the layer of index $i$ is equal to the average distance between two opposite edges of the bubbles; this distance $\ell_i$ is related to the mean bubble area $\mathcal{A}_i$, by  $\ell_i = \sqrt{2\mathcal{A}_i/\sqrt{3}}$.  Note that, due to bubble rearrangements, a layer is not always composed of the same bubbles. We plot in the inset of Fig.~\ref{Fig:areas:fit} the time evolution of the average size $\ell_i$ of the bubbles as a function of their topological distance. Consistently with the snapshots of Figs.~\ref{img:evolution:foam}a and b, this graph shows that the bubble size decreases with time. At a given time, the size also decreases with decreasing topological distance, and the bubbles located deeper in the foam start shrinking only after a delay which increases with the topological distance. At long time, the size of the bubbles closest to the front plateaus at a value about 5 times smaller than the initial size $\ell_0$.

The evolution of the system is driven by the exchanges between the air from the outer atmosphere and the CO$_2$ initially contained in the foam: these gases undergo crossed exchanges to tend to balance their concentrations, a characteristic that differentiates our work from the studies carried out until now on the ripening of bubbles in 2D geometry \cite{Glazier1987,Duplat2011,Roth2013,Saulnier2015,Tong2017,Taccoen2019,Forel2019}. The shrinkage observed here is a direct consequence of the fact that the permeability of CO$_2$ through the soap films of the 2D foam, $k_{\mathrm{CO}_2}$, is larger than that of air, $k_{\mathrm{air}}$ \cite{Farajzadeh2008}: hence, CO$_2$ exits the foam faster than air is incorporated; this behavior is similar to that observed for isolated CO$_2$ bubbles in a liquid at equilibrium with the atmosphere \cite{Shim2014,Penas-Lopez2017}. We recall that in this 2D geometry, the capillary overpressure in the bubbles is of the order the ratio of the liquid-gas surface tension by the gap size, that is a few tens of Pa, and safely negligible compared to the osmotic pressure of the order of 10$^5$ Pa. Since there is no mechanical pressure gradient to drive a net convective flow for the gas mixture, we then expect the dynamics to be diffusive. Accordingly, we rescale the data by using the dimensionless  lengths $\bar{\ell_i} = \ell_i/\ell_0$ and by combining time and topological distance in a single self-similar variable $\sqrt{t}/i$. Fig.~\ref{Fig:areas:fit} shows that plotting  $\bar{\ell_i}$ as a function of $\sqrt{t}/i$ enables to collapse all data on a single master curve, which is indeed the hallmark of a diffusive process \cite{Carslaw2011,Crank1975}.

Even though the dynamics is diffusive, it differs from the usual linear diffusion occurring in an unbounded gas mixture, because each soap film represents a barrier for the gas transfer. To model our observations, we assume that gas transport is limited by the transfer across soap films, and that the gas concentration is uniform within each bubble, and quantified by the molar fraction $x$ of CO$_2$. To simplify the analysis, we neglect the exchanges between neighboring bubbles in the same layer $i$, and we assume that the molar fraction is only a function of time and topological distance, that we denote $x_i(t)$. Under these hypotheses, the volumetric flux of CO$_2$ per unit area from layer $i$ to layer $i + 1$ equals $j_{\mathrm{CO}_2} = -k_{\mathrm{CO}_2} (x_{i+1} - x_i)$ \cite{Princen1965}. Identifying air as a single effective gas of molar fraction $1 - x_i$, its flux per unit area from layer $i$ to layer $i + 1$ equals $j_{\mathrm{air}} = k_{\mathrm{air}} (x_{i+1} - x_i)$. These fluxes dictate both the evolution of the gas contents in each layer, and of the  layer thickness defined as the  size $\ell_i$ of the bubbles in this layer, also assumed uniform within each layer. This ``layered'' model has been previously used to describe the evolution of a train of bubbles inside a tube \cite{Dollet2023}, and yields the coupled evolution equations:
\begin{equation} \label{Eq:discrete_model}
    \begin{split}
    \frac{\mathrm{d}x_i}{\mathrm{d}t} &= -\frac{1}{\ell_i} [k_{\mathrm{air}} x_i + k_{\mathrm{CO}_2}(1 - x_i)] (-x_{i-1} + 2x_i - x_{i+1}) , \\ \frac{\mathrm{d}\ell_i}{\mathrm{d}t} &= (k_{\mathrm{air}} - k_{\mathrm{CO}_2}) (-x_{i-1} + 2x_i - x_{i+1}) .
    \end{split}
\end{equation}
The initial conditions are $x_i = x^0$ (if the foam is initially filled with CO$_2$  then $x_i = 1$) and $\ell_i = \ell_0 $ at $t = 0$. Combining the two Eqs. \ref{Eq:discrete_model} shows at once that $\mathrm{d}\{ [k_{\mathrm{air}} x_i + k_{\mathrm{CO}_2}(1 - x_i)]\ell_i \}/\mathrm{d}t = 0$. Hence, from the initial condition $x_i=x^0$ in the foam, the evolutions of the bubble size and of the molar fractions are directly correlated: \begin{equation}\label{Eq:l_i}
\ell_i = \frac{k_{\mathrm{air}} x^0+k_{\mathrm{CO}_2}(1 - x^0)}{k_{\mathrm{air}} x_i + k_{\mathrm{CO}_2}(1 - x_i)} \ell_0.
\end{equation}

This equation, which relates $x_i$, the local molar fraction of CO$_2$, to $\ell_i$, the characteristic bubble size, is particularly useful for 2D foams where the evolution of $\ell_i$ can be easily captured  with a simple camera.
Inserting Eq. \ref{Eq:l_i} in Eq. \ref{Eq:discrete_model} shows that it suffices to solve the following evolution equation:
\begin{equation} \label{Eq:discrete_model_xi}
   \frac{\mathrm{d}x_i}{\mathrm{d}t} = -\frac{ [k_{\mathrm{air}} x_i + k_{\mathrm{CO}_2}(1 - x_i)]^2 (-x_{i-1} + 2x_i - x_{i+1}) }{[k_{\mathrm{air}} x^0+k_{\mathrm{CO}_2}(1 - x^0)] \ell_0}.
\end{equation}

Instead of dealing with a discrete set of equations, we assume that the number of layers is large enough to treat $i$ as a continuous variable, and define the functions $x(i,t)=x_i(t)$ and $\ell(i,t)=\ell_i(t)$. Accordingly, we identify the finite difference $x_{i+1} - x_i$ to a (continuous) gradient $\partial x/\partial i$. Hence, the term $-x_{i-1} + 2x_i - x_{i+1}$ becomes $-\partial^2 x/\partial i^2$. Moreover, we can make this problem dimensionless by setting $\bar{\ell}= \ell/\ell_0$ and $\bar{t} = t/\tau$ with the characteristic time $\tau = \ell_0/k_{\mathrm{CO}_2}$. With these transformations, the set  of discrete evolution equations (Eqs. \ref{Eq:discrete_model_xi}) becomes a dimensionless partial differential equation:
\begin{equation} \label{Eq:continuous_model}
   \frac{\partial x}{\partial \bar{t}} = \frac{(1-x + \varepsilon x )^2}{1-x^0+\varepsilon x^0}  \frac{\partial^2 x}{\partial i^2},
\end{equation}
with $\varepsilon = k_{\mathrm{air}}/k_{\mathrm{CO}_2}$ the permeability ratio. The nonlinear character of the diffusion process is highlighted by the presence of the initial conditions in this evolution equation. In our experimental conditions, $x$ obeys the initial condition $x^0=x(i>0,\bar{t} = 0) =  1$, and the boundary conditions $x(i = 0,\bar{t}) = 0$ (the front connects foam and atmosphere whose molar fraction of CO$_2$, equal to $4\times 10^{-4}$, can safely be neglected here) and $x(i\rightarrow\infty,\bar{t}) \rightarrow 1$ (there is no evolution in the foam far away from the front).

This mathematical problem has the structure of a diffusion equation, but with a nonlinear term coming from the evolution of the gas mixture inside each layer. As usual with diffusion problems without external length, it admits a self-similar solution of the form $x(i,\bar{t}) = x(\xi)$ with $\xi = i/\sqrt{\bar{t}}$, which obeys, when $x^0=1$, the following ordinary differential equation:
\begin{equation} \label{Eq:ODE_x_xi}
    -\xi \frac{\mathrm{d}x}{\mathrm{d}\xi} = \frac{2}{\varepsilon} (1-x + \varepsilon x)^2 \frac{\mathrm{d}^2 x}{\mathrm{d}\xi^2} ,
\end{equation}
with boundary conditions $x(\xi = 0) = 0$ and $x(\xi\rightarrow\infty) \rightarrow 1$. 

The self-similar character of the evolution is fully supported by the excellent collapse of the experimental data (Fig.~\ref{Fig:areas:fit}) when using the self-similar variable $\xi$. We can go further and fit the experimental master curve by the numerical solution of Eq. \ref{Eq:ODE_x_xi} with the two permeabilities as fitting parameters. This  procedure also yields an excellent agreement with the data, fully quantitative over all the range of the rescaled variable (Fig.~\ref{Fig:areas:fit}); it yields the values of the two permeabilities as fitting parameters $k_{\mathrm{air}} = (1.3 \pm 0.2) \times 10^{-4}$~m/s and $k_{\mathrm{CO}_2} = (6.0 \pm 0.7) \times 10^{-4}$~m/s, which we shall discuss later. Notice that our model gives a straightforward interpretation of the long-time plateau, characterized by $\ell/\ell_0 \to \varepsilon$.  According to Eq. \ref{Eq:l_i}, it corresponds to $x=0$, i.e. when air has fully invaded the foam. Therefore, a measurement of the final state only directly yields the permeability ratio of the two gases. The presented experiment was reproduced five times (see Materials and Methods), confirming all the aforementioned results, with very little quantitative differences (the standard deviations appearing in the values of $k_{\mathrm{air}}$ and $k_{\mathrm{CO}_2}$ come from the dispersion between different experiments). In summary, our diffusive model seems to perfectly capture the dynamics of transfer of a mixture of gas in a liquid foam.

\subsection*{Determination of an effective diffusivity} To go one step further and  extract effective transfer properties at the whole foam scale in view of large-scale sizing of separation setups, we switch back to the real space variables and discuss the evolution of gas molar fraction at distance $Z$ from the front line. We introduce the rescaled distance $\bar{Z}=Z/\ell_0$ and the self-similar variable $\zeta=\bar{Z}/\sqrt{\bar{t}}$, which is related to $i$, the number of the layer located at distance $\bar{Z}$, by $\zeta=\int_0^{\xi}[\ell(u)/\ell_0] \mathrm{d}u\equiv F(\xi)$.

\begin{figure}
    \centering
        {\includegraphics[width=0.95\linewidth]{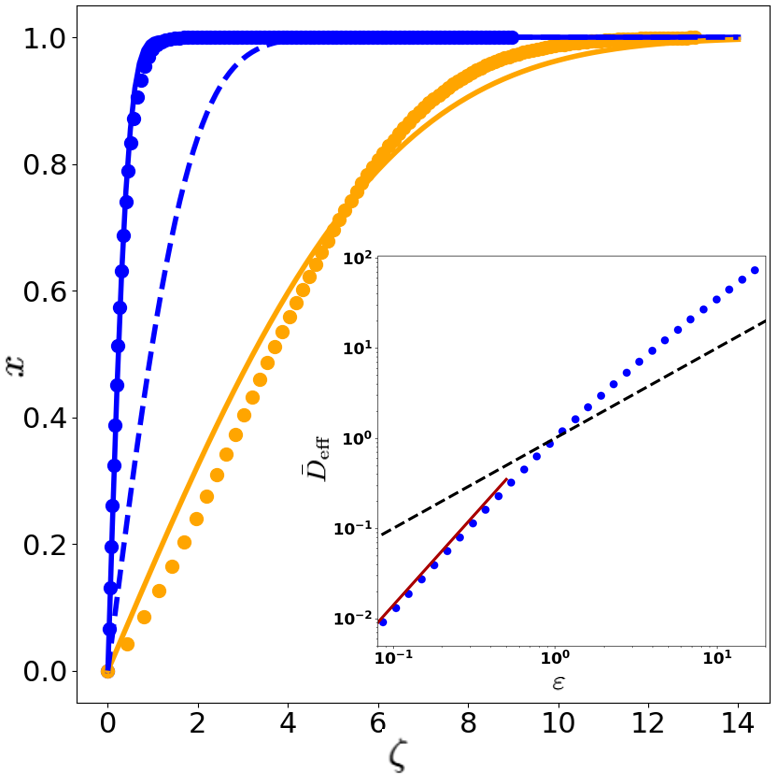}}

    \caption{Blue dots: plot of molar fraction $x$ as a function of the self similar variable $\zeta$, as found from our non-linear diffusion model (Eq. \ref{Eq:ODE_x_xi} and initial conditions $x=0$ at the front and $x=1$ initially in the foam), with $\varepsilon=0.22$, and fit with the solution of the diffusion equation \ref{eq:diffstandard} with boundary conditions $x(0)=0$ and $x(\zeta\to\infty)=1$. This provides an effective diffusion coefficient $\bar{D}_{\mathrm{eff}}=0.06$ in reduced units. Data in orange are similarly calculated, with $\varepsilon=1/0.22\simeq 4.5$. They yield $\bar{D}_{\mathrm{eff}}=11.4$. Dashed line show the theoretical solution for $\bar{D}_{\mathrm{eff}}=1$, as a guideline. Inset shows $\bar{D}_{\mathrm{eff}}$ as a function of $\varepsilon$ in log-log scale: blue dots are approximate values found by the fits, when $x=0$ at the front and $x=1$ initially in the foam; dashed line indicates the limit case $\bar{D}_{\mathrm{eff}}=\varepsilon$ corresponding to vanishing initial concentration gradient. The red curve is the prediction $\bar{D}_{\mathrm{eff}} = 1.40\varepsilon^2$ from the asymptotic analysis at $\varepsilon\ll 1$ (see Materials and Methods). 
    } \label{Fig:Deff}
\end{figure}

In order to gain genericity, we assume that our modeling is valid for all values of $\varepsilon$ though we validated it only for $\varepsilon=0.22$. Using the numerical solution of Eq. \ref{Eq:ODE_x_xi}, we plot in Fig. \ref{Fig:Deff} the molar fraction of the gas, which was initially in the foam, $x\big(F^{-1}(\zeta)\big)$, as a function of $\zeta$, for $\varepsilon=0.22$, 1 and $1/0.22$.

To obtain an effective diffusivity, we compare these data with the solution of the classical diffusion equation in a semi-infinite space, introducing $\bar{D}_{\mathrm{eff}}$, a constant dimensionless diffusion coefficient:  
\begin{equation}   \frac{\partial x}{\partial \bar{t}} =\bar{D}_{\mathrm{eff}}\frac{\partial^2 x}{\partial \bar{Z}^2} \label{eq:diffstandard}\end{equation} 
Taking $x(\bar{Z}=0,\bar{t}) = 0$ and $x(\bar{Z}>0,\bar{t}=0) =1$ for boundary and initial conditions, Eq. \ref{eq:diffstandard} leads to the solution $x_{\mathrm{hom}}(\zeta)=\erf\big(\zeta/2 \sqrt{ \bar{D}_{\mathrm{eff}}}\big)$, where $\erf$ is the error function.

When  $\varepsilon=1$, the flux of the two gases are equal in intensity but with opposite direction, thus the bubble size remains constant and the foam does not evolve, yielding $\ell=\ell_0$ at all times therefore $\bar{Z}=i$. Eq.~\ref{Eq:continuous_model} is then identical to Eq.~\ref{eq:diffstandard} when $\bar{D}_\mathrm{eff}=1$. In real-space units, $D_\mathrm{eff}$, the diffusion coefficient is then $D_\mathrm{eff}=D_0=\ell_0 k_{\mathrm{CO}_2}$, i.e. the two gases diffuse identically by crossing a barrier of permeability $k_{\mathrm{air}}=k_{\mathrm{CO}_2}$ every (fixed) distance $\ell_0$, as discussed for a three-dimensional foam in \cite{Trinh2023}.

For our test case of $\varepsilon=0.22$, Fig. \ref{Fig:Deff} shows that $x(\zeta)$ is rather well fitted by $x_{\mathrm{hom}}(\zeta)$ with $\bar{D}_\mathrm{eff}=0.06$: the foam can therefore be considered as an homogeneous material, notwithstanding its heterogeneous structure as a foam. By contrast, if we consider $\varepsilon=0.22^{-1}\simeq 4.5$, corresponding in particular to a foam of air in contact with a CO$_2$ atmosphere, $\bar{D}_\mathrm{eff}$ is equal to 11, but in that case, describing the diffusive process by a linear model is not as accurate as for $\varepsilon<1$, as indicated by the approximate fit in Fig. \ref{Fig:Deff} in the $\varepsilon>1$ limit.

To understand the strong (superlinear) increase of $\bar{D}_\mathrm{eff}$ with $\varepsilon$, we come back to the nonlinear Eq. \ref{Eq:ODE_x_xi} which can be seen as a classical diffusion equation with a local diffusion coefficient $ (1-x + \varepsilon x) ^2/\varepsilon$. At time $t = 0$ when $x=1$, this coefficient is thus equal to $\varepsilon$. In the extreme case where $\varepsilon=0$ (i. e. air does not  cross the liquid films), the dynamics is even stuck: even though CO$_2$ from the first layer escapes across the front line, all layers keep $x=1$ since there is no air entering into them, therefore there will be no flux between the different layers inside the foam and $\bar{D}_\mathrm{eff}=0$. To understand the diffusive process in the real space, we also need to recall that the real space variable $Z$ is related to the layer numbers of Eq.  \ref{Eq:continuous_model} by factors $\ell$, that converge to $\varepsilon\ell_0$ at short distance or long times. This $\varepsilon$ correction factor in the distances induces a $\varepsilon^2$ correction factor in the effective diffusion coefficient. For small values of $\varepsilon$, it is actually possible to go beyond this scaling argument: a straightforward asymptotic analysis of the nonlinear model is provided in Materials and Methods, yielding the prediction $\bar{D}_\mathrm{eff} = 1.40\varepsilon^2$, in excellent agreement with the data (see inset of Fig.~\ref{Fig:Deff}). Thus, whether at short times, near the front or at long times, we observe a strong dependence of $D_\mathrm{eff}$ on $\varepsilon$, suggesting nontrivial variation in the whole time-space domain as depicted in the inset of Fig.~\ref{Fig:Deff}, where $D_\mathrm{eff}$ is observed to increase with $\varepsilon^\alpha$ with $\alpha$ between 1.5 and 2. 

Another interesting case is that of small initial gradient, \textit{i.e.} $x$ close to 1 near the front. To the leading order, the local diffusion constant of Eq. \ref{Eq:ODE_x_xi} becomes global and equal to $\varepsilon$. Together with the previous case, this provides a range for the expected diffusivities for a foam filled with a given gas in contact with an atmosphere where this gas is diluted in an unknown ratio.

Back to our CO$_2$/air case, we therefore notice that in the case of a CO$_2$-laden foam, the effective diffusion coefficient should be between $D_{\min}=0.06\, k_{\mathrm{CO}_2} \ell_0 \approx  2.5\times 10^{-7}$~m$^2$/s, as in our experimental case, and $D_{\max}= \varepsilon k_{\mathrm{CO}_2} \ell_0= k_{\mathrm{air}} \ell_0 \approx 9\times 10^{-7}$~m$^2$/s, in case of a tiny gradient between the foam and the front. 

Considering an air-laden foam amounts to considering the case $\varepsilon=k_{\mathrm{CO}_2}/k_{\mathrm{air}}$, while keeping in mind that we normalised the permeabilities by that of the gas initially trapped in the foam. We then find a lower bound $D_{\min}=\varepsilon k_{\mathrm{air}} \ell_0= k_{\mathrm{CO}_2} \ell_0 \approx 4\times 10^{-6}$~m$^2$/s and an upper bound $D_{\max}\approx  11\,k_{\mathrm{air}} \ell_0 \approx  1\times 10^{-5}$~m$^2$/s.

 Note that while in the  case of CO$_2$-laden foam the diffusion coefficient is intermediate between the typical diffusivities of gas in water, in the order of $10^{-9}$~m$^2$/s and gas in another gas, in the order of $10^{-5}$~m$^2$/s, in the second case it is close to the diffusion coefficient in gas, a consequence of the bubble growth leaving less films to cross per unit length. In this latter case, however, the assumption of instantaneous diffusion within each bubble needs to be reconsidered.

\subsection*{Transport of bubbles}

\begin{figure}
    \centering
        {\includegraphics[width=\linewidth]{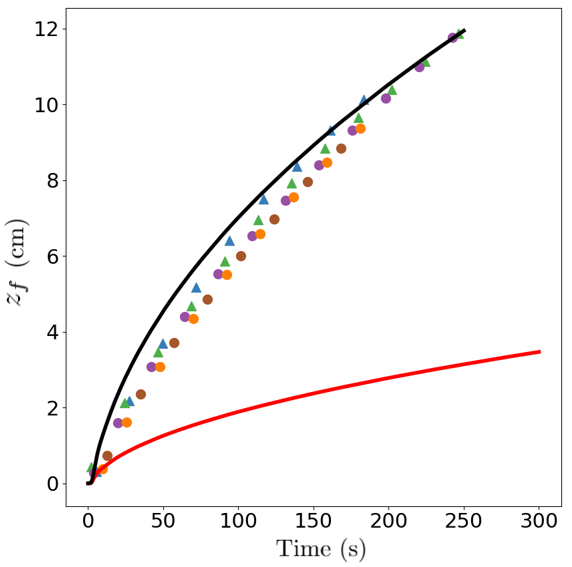}}

    \caption{Time evolution of the mean distance of the front with respect to the open end of the cell $z_f(t)$ for five experimental realizations (distinguished by different symobls and colors). The red curve corresponds to the prediction of Eq. \ref{Eq:frontnaif}. The black curve corresponds to the prediction of Eq. \ref{eq:zfpur}.} \label{Fig:front}
\end{figure}

While the one-dimensional layered model accounts well for the diffusive processes of the gas, with the layer index as reference space coordinate, we now turn to the evolution of the structure of the foam. An important observable in our experiments is the dynamics of the front. Fig.~\ref{img:evolution:foam}a shows that it gets corrugated as soon as its starts retracting in the cell. The corrugations vary between different experiments. However, Fig.~\ref{img:evolution:foam}b also shows the amplitude of these corrugations remains relatively small with respect to the position $z_f(t)$ of the front, defined as its mean distance with respect to the open end of the cell. The quantity $z_f(t)$ is plotted in Fig.~\ref{Fig:front} for five experimental realizations, showing that the front progresses inside the cell, with a speed that decreases as a function of time; there is also a good reproducibility between different experiments, despite the aforementioned variabilities in the front shape.

The front position is related to  $S(t)$ the area covered by the bubbles by $z_f(t)=(S(0)-S(t))/w$. Since the front dynamics is driven by the shrinkage of the bubbles, and the model predicts the thickness of each layer $\ell_i(t)$, it seems natural to predict $z_f(t)$ from the cumulative decrease of thickness of each layer: 
\begin{equation} z_f(t) = \sum_{i=0}^\infty [\ell_0 - \ell_i(t)],\label{Eq:frontnaif} \end{equation}
This prediction clearly underestimates the front displacement (Fig.~\ref{Fig:front}).

To explain this discrepancy, it is necessary to consider the two-dimensional plastic rearrangements of the bubbles. Indeed, as the bubbles shrink, more of them are required to fill the layers of fixed width between the two side walls of the Hele-Shaw cell, since bubbles in foams usually do not deform by more than a yield strain of order 10$\%$ \cite{DolletRaufaste}. Consequently, the foam experiences rearrangements, with bubbles from ``deeper'' layers being recruited to fill gaps between shrinking bubbles, as illustrated in Fig.~\ref{img:evolution:foam}c. To bring these bubbles back from the deep layers to the front, we can envisage either a displacement from the deep layers to the front, or from the front to the deep layers. Here, moving from the deep layers to the front involves moving a large number of liquid menisci, and therefore high friction, and is not observed in practice. Thus, the front is more mobile than the deep layers, and the recruitment of bubbles involves an additional motion of the front towards the deep layers.

A consequence of the ability of the layer model to describe the size of bubbles in spite of the many bubble reorganizations is that we can consider that a bubble reaching layer $i$ instantaneously adopts an area $\mathcal{A}_i(t)=\ell_i(t)^2 \sqrt{3}/2$ that is set by the layer width $\ell_i(t)$. Note that this can be achieved even without any motion of the bubble, since a plastic event in other layers closer to the front may directly affect its topological distance to the front. The number $N_i(t)$ of bubbles that fills in the layer $i$ is given by $N_i=w \ell_i/\mathcal{A}_i=N_0 \ell_0/\ell_i$, where $N_0=w \ell_0/ \mathcal{A}_0$ is the initial number of bubbles in each layer. Comparing the actual area of these bubbles with their initial one, we obtain
\begin{equation}
z_f(t)=\frac{1}{w}\sum_{i=1}^\infty [\mathcal{A}_0-\mathcal{A}_i(t)] N_i(t) =\ell_0\sum_{i=1}^\infty \left[ 1-\frac{\ell^2_i(t)}{\ell^2_0} \right] \frac{\ell_0}{\ell_i(t)}. \label{eq:zfpur} \end{equation}

As shown in Fig. \ref{Fig:front}, the proposed expression for $z_f$, where we have simply used the expression for $\ell$ obtained using the fit of the experimental data with the numerical solution of Eq. \ref{Eq:ODE_x_xi}, provides an excellent agreement with the experimental measurements of the front position, with no additional fitting parameter. 

We go one step further in the comprehension of the evolution of the foam structure by examining the flux of bubbles from layers to layers. A characterization of this diffusion-induced advection may turn out  useful in case the foam also carries non (or slowly) diffusing species. 

\begin{figure}
    \centering
        {\includegraphics[width=0.95\linewidth]{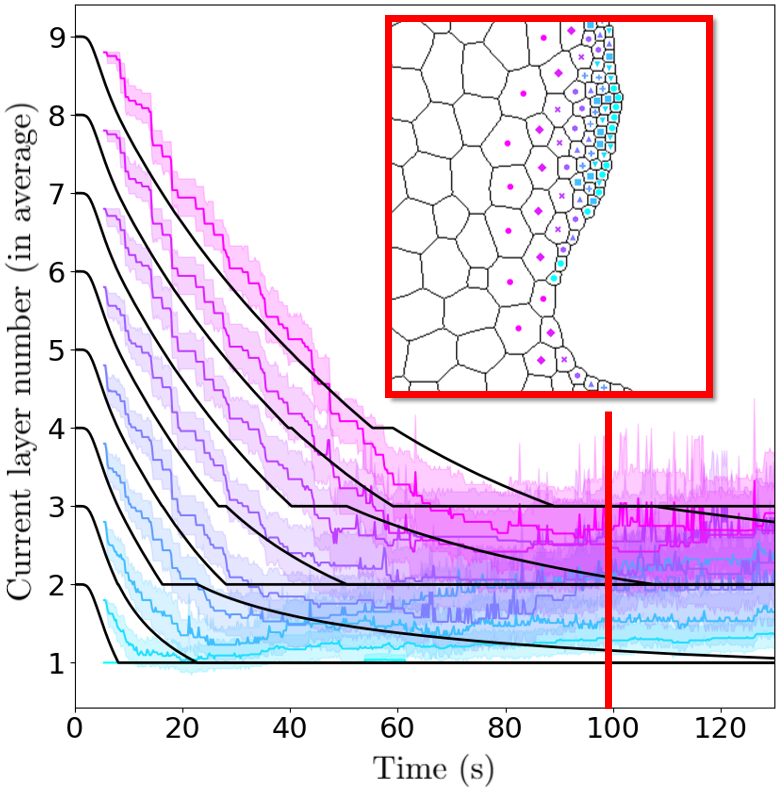}}
    \caption{\label{img:evolution_topological_distance} Evolution of the average current topological distance $I_i$ of the bubbles initially located on layer $i$ ; the eight curves correspond to the  different initial topological distances 2 to 9, as indicated by the color code. The black curves are the prediction of the model (Eq. \ref{eq:Ii}). Note that as the bubbles can only be followed after $t_0\simeq 9$ s (see Materials and Methods), the initial value of $I_i(t_0)$ is not known and is initially set as $i$; then all the curves were shifted downwards by the same amount $\simeq 0.2$ such that $\min[I_2(t)]=1$, a reasonable assumption. Inset: snapshot of the foam at time $t = 99$~s, reproducing the boxed area of Fig.~\ref{img:evolution:foam}b. The bubbles are marked according to their initial topological distance (same symbols as Fig.~\ref{img:evolution:foam}a). This illustrates the fact that the final topological distance is lower than the initial one.}
\end{figure}

Having tracked all bubbles (see Materials and Methods), we present in Fig.~\ref{img:evolution_topological_distance} the time evolution of $I_i(t)$, defined as the average current topological distance of the bubbles initially at a topological distance $i$. All curves display the same evolution: first, the current topological distance decreases, which is a signature that some bubbles move from one layer to the next towards the front; then, the curves saturate. Both the transition time between these two behaviors and the saturation value increase with the initial topological distance. More precisely, in the early times when $I_2$ gets closer to 1 (i.e. bubbles from layer 2 get in contact with the front), all other curves decrease with a similar slope; this indicates that most motions occur between layer 2 and 1 and that the decrease of $I_i, i\ge 3$ is the sole mathematical consequence of the loss of topological distance that is transmitted neighbor by neighbor. When the $I_2$ curves saturates, the process continues with the next layers. We can then assume a sequential evolution of the bubbles, i.e. bubbles of layer $i$ having reached layer $k$ will only move to layer $k-1$ (if needed) when all bubbles initially in layer $j$, $k-1<j < i$ will already have done so. Under this assumption, and after some rigorous counting (see Materials and Methods),  we can estimate $\Theta^+_{i,k}(t)$, defined as the fraction of bubbles initially in layer $i$ that have at least moved to layer $k$ at time $t$ as \begin{equation}
 \Theta^+_{i,k}(t)   =\min\left\{ 1,\max\left[ 0,1-i+\sum_{j=1}^k \frac{\ell_0}{\ell_j(t)} \right]\right\}, \label{eq:flux} \end{equation}
allowing to estimate $I_i$ as
\begin{equation}\label{eq:Ii} I_{i}(t)= i-\sum_{k=1}^{i-1} \Theta^+_{i,k}(t).\end{equation}

As for the front position, this expression only involves the diffusion-controlled function $\ell_i(t)$, for which we can consider the fitting function of the experimental data shown in Fig. \ref{Fig:areas:fit}. As shown in Fig. \ref{img:evolution_topological_distance}, our geometry-based estimate of the bubble flux averaged over the cell width accounts well for the observation.

\section*{Discussion}
\subsection*{Permeabilities of air and CO$_2$} Although the measurement of gas permeabilities through a soap film is not the main goal of our study, the permeabilities are fitting parameters of our model, that can be compared with the results of the literature. Gas permeability is generally measured studying unsteady gas transfers across soap films. Gas transfer occurs when the soap film separates two gas mixtures, (i) either of same chemical composition but one slightly overpressurized, as in the diminishing bubble method \cite{Princen1965,Tcholakova2011} and in coarsening of 2D foams \cite{Saulnier2015, Forel2019}, (ii) or of the same pressure but with a difference in chemical composition (N$_2$ on one side and O$_2$ on the other, for example) \cite{Cook1974,Hadji2020,Hadji2024}. Since the permeability of a soap film depends on the ability of gas molecules to cross surfactant barriers as well as to solubilize and diffuse through the aqueous core of the film, comparisons are often difficult to make \cite{Farajzadeh2008}. Two specific cases stand out, however. Firstly, the absolute value of permeabilities measured here can be compared with results from experiments where the thickness of the soap film has reached the smallest possible value, in the order of tens of nanometers \cite{Farajzadeh2011}; As permeability decreases with film thickness, the permeability through these ultra-thin films, called Newton black films (NBFs), gives an upper bound for our permeability value.

The permeability of N$_2$,  O$_2$ and air through SDS Newton-black films, have been found to be $k_{\mathrm{N}_2} \sim 1.5\times 10^{-3}$~m/s, $k_{\mathrm{O}_2} \sim 3 \times 10^{-3}$~m/s and $k_{\mathrm{air}} \sim 2 \times 10^{-3}$~m/s respectively  \cite{Ramanathan2011, Forel2019}. There are far fewer data available for CO$_2$, as transfer kinetics through soap films are much faster due to the large solubility of CO$_2$ in the aqueous core of the films, and to the best of our knowledge there are no data for SDS NBFs \cite{Farajzadeh2011}. Nevertheless, Princen and Mason measured permeabilities value for thin films stabilized with cationic surfactant for different gases and found $k_{\mathrm{CO}_2} = 75.5 \times 10^{-3}$~m/s and $k_{N_2}=1.4\times 10^{-3}$~m/s, the latter value being very close from the one measured for SDS NBFs. Note that the authors stress that for CO$_2$, the transfer kinetics are so rapid that the film may not be a NBF \cite{Princen1965}.  
Overall, these values are one to two orders of magnitude larger than ours. This suggests that our foam films are quite thick, with thicknesses between 100 nm and 1 µm. There are several reasons for this. Firstly, the films through which permeabilities are measured are moving lamellae sliding along the glass walls at speeds of the order of mm/s \cite{Hadji2024}. This dynamic character can significantly increase film thicknesses up to 100 nm \cite{Emile2007}. In addition, the rearrangements that accompany bubble shrinkage create new films, thicker than NBFs. Thus, our permeability values would be lower than those reported in the literature for NBFs because our films are not at equilibrium \cite{Tregouet2023}.

Then, the bubbles in our experiments are separated not only by thin films, but also by liquid channels (so-called Plateau borders) either between three bubbles, or between two bubbles and a wall. These liquid channels are orders of magnitude thicker than soap films: they have a radius of curvature $R_{PB}$ equal to 0.35~mm in our experiments (see Fig.~\ref{img:evolution:foam}c). Hence, gas transfer is negligible through liquid channels and occurs through the thin film regions only \cite{Schimming2017}. Accordingly, we should correct our measured values of permeabilities by the relative proportion of the boundaries between bubbles occupied by thin films \cite{Hilgenfeldt2001,Forel2019}. Last, we assumed that the gas concentration in each bubble was uniform, and that the foam is always in contact with pure air at the front. These two hypotheses amount to neglecting the interdiffusion of air and CO$_2$ in each connected region (the bubbles, or the atmosphere entering the cell as the foam retracts) or, equivalently, to assuming that the diffusion coefficient in gas phase $D_g = 1.9\times 10^{-5}$~m$^2$/s is much larger than the effective diffusivity $D_{\mathrm{eff}} \approx  0.06\, k_{\mathrm{CO}_2} \ell_0$ previously introduced. With $k_{\mathrm{CO}_2} = 6\times 10^{-4}$~m/s and $\ell_0 = 0.70$~cm, we evaluate $D_{\mathrm{eff}} = 2.5\times 10^{-7}$~m$^2$/s, which is two orders of magnitude lower than $D_g$. In this configuration, our hypothesis then seems to be valid.

The second point worth discussing is  the value found for $\varepsilon$, the ratio between the permeabilities of air and CO$_2$, which is 0.22 in our experiment.  If we assume that the solubilities and diffusivities of these two gases in water are unaffected by the presence of surfactants, then using the value from literature for solubilities and diffusivities, the soap film permeability ratio must be equal to 0.020, a value close to the experimental value of $0.016$ obtained by dividing the permeabilities measured by Princen and Mason for pure CO$_2$ and then for pure N$_2$ \cite{Princen1965} and $0.021$ measured by Cook and Tock by connecting a single soap film to two atmospheres containing CO$_2$ and N$_2$ at different concentrations \cite{Cook1974}. This is very different from $\varepsilon=0.22$ measured here and from the results of Hadji \textit{et al.} who measured a ratio between $0.1$ and $0.22$ using the same set-up than Cook and Tock but for CO$_2$ and air. The only difference between the different works is the nature of the gases used: Cook and Tock used binary mixtures of $\mathrm{CO}_2$ and $\mathrm{N}_2$, whereas Hadji uses mixtures of $\mathrm{CO}_2$ and air or $\mathrm{N}_2$ and air. Princen and Mason proved in their pioneering paper on the permeability of soap films that air could be considered an effective gas in the diminishing bubble method experiment \cite{Princen1965}. However, this is only valid in the case of a binary mixture of two gases for transfers with an imposed pressure gradient, and not for ternary mixtures of gases with an imposed osmotic pressure as in our experiments. The presence of O$_2$, a third gas that is more permeable than N$_2$ in soap films, may modify the composition of the first layers of bubbles, limit their deflation and thus increase the value of $\bar{\ell}$ at long time and thus overestimate $\varepsilon$.

\subsection*{Effective diffusivity} We conclude our discussion by focusing on the main output of our study, the effective diffusion constant at the foam scale. We recall that we only consider transfer of a gas trapped within an aqueous foam when the latter is brought into contact with an atmosphere of different composition. The diffusive process considered here is therefore quite different from the one discussed in \cite{Hilgenfeldt2001} on foam coarsening, where gas transfer is induced by capillary pressure gradients between foam bubbles. Indeed, in our work, the amplitude of concentration gradients effects (a.k.a. osmotic effects) scales as  $\Delta x$~$P_0\sim 10^5$, where $\Delta x \sim 1$ represents the initial molar fraction difference between the foam and the atmosphere, and is thus way larger than capillary overpressure (for millimetric bubbles, the latter is at most 10$^2$~Pa). Another fundamental difference lies in the evolution of the bubble size distribution: in capillary driven foam coarsening, the mean size and standard deviation of the distribution increase with time, but the foam remains spatially homogeneous with the smallest bubbles evenly distributed in the foam volume, whereas here the mean bubble size decreases (or increases, depending on the value of $\varepsilon$) and the distribution of small bubbles becomes very heterogeneous with the smallest bubbles located only at the air/foam boundary.

We also emphasize that in this osmotic limit, gas diffusion is highly nonlinear and this has some interesting consequences, which we illustrated through the contrast between the case of a CO$_2$-laden foam in contact with air, and that of a foam full of air in contact with pure CO$_2$, for which the effective diffusion constants differ by a factor 40. The diffusion is thus enhanced when the less mobile gas is initially trapped in the foam, while it would be slowed down when the more mobile gas is in the foam. For a given gas initially trapped in the foam, the inset of Fig. \ref{Fig:Deff} suggests that its diffusion coefficient can vary by several order of magnitudes depending on the composition of the atmosphere. In practice, it would vary between the diffusion coefficient in liquid and that in gas, corresponding respectively to two limiting cases where the model would need to be refined. This is yet four orders of magnitude that can be spanned. Conversely, for a given atmospheric composition, gases of different permeabilities will diffuse differently. This results is particularly striking in as much the diffusion coefficients of most gases are fairly similar both in a gaseous environment (of the order of $10^{-5}$ m$^2$/s) and in a liquid phase (of the order of $10^{-9}$ m$^2$/s). By contrast, solubilities of gases in water vary by several orders of magnitudes, enabling \textit{in fine} active control of diffusive behaviour. This surprising potentiality suggests that aqueous foams could be used in low-cost processes to block and then release a gas on demand by simply modifying the chemical composition of the atmosphere above the foam.

 Additionally, we stress that this diffusion process takes place from the front into the foam, but that the position of this front is itself dependent on the diffusive process. A model for the front position has been proposed, which relies on the hypothesis of instantaneous equilibrium of a bubble size with that of its new layer. This hypothesis is supported by the fact that the local transfer rate, given by $D_0=k_{\mathrm{CO}_2} \ell_0$, is larger than the global one, given by $D_{\mathrm{eff}}$. For other initial distribution of gases, this assumption may thus need to be reconsidered.

We may wonder if our results for two gases can be extrapolated to three-gas mixtures that would bring one additional control parameter. What happens if two gases of different permeability are stored above an atmosphere composed of a third gas? A study on the scale of a single bubble laden with CO$_2$ and CH$_4$ in contact with a N$_2$ atmosphere indeed showed that the more mobile gas (CO$_2$) diffuses rapidly in the atmosphere while the less mobile one (CH$_4$) remains trapped in the bubble \cite{Lin2022}. The effective diffusion coefficients of the three gases in this process should now be measured to quantify the magnitude of the coupling between gas transfer and bubble structure, as in the present work. Although this work remains to be done, these results taken together suggest that assembly of soap films could well be potential game changers for low-tech membrane separation processes.

\section*{Materials and Methods}

\subsection*{Experiments}We use a Hele-Shaw cell of length 70~cm, width $w = 15$~cm and gap $h = 2$~mm. The downstream end of the cell is open to the atmosphere. We fill a reservoir located upstream the cell with a solution of sodium dodecyl sulfate (SDS) of concentration 10~g/L. We generate a foam by bubbling CO$_2$ at a controlled rate of $Q_g=30$~mL/min through a nozzle at the bottom of the reservoir.
Some additional solution is also injected in the reservoir using a syringe pump at rate $Q_\ell$. The resulting foam forms a bubble monolayer in the cell.

The experiment has been repeated five times to assess its reproducibility, with the same bubbling rate $Q_g$. We have varied $Q_\ell$ between 60 and 120~mL/h, and found that this parameter has no significant influence of the measurements. For each experiment, the initial average area $\mathcal{A}_0$ of the bubbles is measured by image analysis on the first image ; its value is $\mathcal{A}_0 = 0.43 \pm 0.02$~cm$^2$, with a standard deviation coming from the difference of initial bubble area of the five experiments. We also measured manually the radius of the Plateau borders between two neighboring bubbles and the wall (Fig.~\ref{img:evolution:foam}c). Averaging over a few tens of Plateau borders yields $R_{PB} = 0.35 \pm 0.01$~mm.

For each experiment, we record a series of $2 \times 10^{3}$ images with a frame rate of 8.99 frames per second. The images are then binarized and skeletonized. Each bubble is labelled on the first image, then tracked over time; the tracking is made by detecting closest bubble centroids between two consecutive images. Any new spurious bubble (such as a Plateau border wrongly detected as a bubble) is discarded to minimize false detection. We measure the time evolution of the area of each bubble, and of its instanteneous topological distance $i$, defined as the minimal number of films separating the considered bubble to the front.

\subsection*{Asymptotic analysis of the model}

We perform an asymptotic analysis of Eq.~[\ref{Eq:ODE_x_xi}] in the limit $\varepsilon\ll 1$, based on matched asymptotic expansions, to derive the asymptotic value of the effective diffusivity in this limit.

We start setting $X = 1 - x + \varepsilon x$. Eq.~[\ref{Eq:ODE_x_xi}] can then be recast as:
\begin{equation} \label{Eq:ODE_X_xi}
    -\xi \frac{\mathrm{d}X}{\mathrm{d}\xi} = \frac{2}{\varepsilon} X^2 \frac{\mathrm{d}^2X}{\mathrm{d}\xi^2} ,
\end{equation}
and the boundary conditions become $X(0) = 1$ and $X(\xi\rightarrow\infty) \rightarrow \varepsilon$.

The boundary conditions suggest two different outer expansions for this problem: at large $\xi$, we set $X = \varepsilon\tilde{X}$. Expanding [\ref{Eq:ODE_X_xi}] yields at leading order $\mathrm{d}\tilde{X}/\mathrm{d}\xi = 0$, with boundary condition $\tilde{X}(\xi\rightarrow\infty) \rightarrow 1$. This yields the trivial solution $\tilde{X} = 1$, obviously incompatible with the boundary condition at $\xi = 0$, which suggests another outer expansion at small $\xi$, of the form $X = \hat{X}$. Expanding [\ref{Eq:ODE_X_xi}] yields at leading order $\mathrm{d}^2 \hat{X}/\mathrm{d}\xi^2 = 0$, with boundary condition $\hat{X}(0) = 1$. The corresponding solution is $\hat{X} = 1 - \xi/A$ with a (yet) undetermined constant $A$, which we anticipate to be strictly positive for matching with the large-$\xi$ solution to be possible. This small-$\xi$ outer solution loses its validity when it becomes of order $\varepsilon$, hence when $A - \xi$ becomes of order $\varepsilon$.

The breakdown of both outer expansions suggests the existence of a transition layer around $\xi = A$, over a length of order $\varepsilon$, where the solution is of order $\varepsilon$. Hence, we seek an inner solution of the form $X = \varepsilon\bar{X}(t)$, with $\xi = A + \varepsilon t$. Substituting in [\ref{Eq:ODE_X_xi}] yields at leading order:
\begin{equation} \label{Eq:ODE_X_t}
    -A\frac{\mathrm{d}\bar{X}}{\mathrm{d}t} = 2\bar{X}^2 \frac{\mathrm{d}^2 \bar{X}}{\mathrm{d}t^2} ,
\end{equation}
with matching conditions with both outer solutions: $\bar{X}(t\rightarrow +\infty) \rightarrow 1$ on one hand, and $\bar{X}(t\rightarrow -\infty) \rightarrow +\infty$ and $\mathrm{d}\bar{X}/\mathrm{d}t (t\rightarrow -\infty) \rightarrow -1/A$ on the other hand. Integrating Eq.~[\ref{Eq:ODE_X_t}] once yields $2\mathrm{d}\bar{X}/\mathrm{d}t = A/\bar{X} + B$ with $B$ an integration constant. Applying the matching condition at $t\rightarrow -\infty$ gives $B = -2/A$. Now, the matching condition at $t\rightarrow +\infty$ shows that $\mathrm{d}\bar{X}/\mathrm{d}t$ must equal zero for $\bar{X} = 1$. This imposes that $A - 2/A = 0$, hence $A = \sqrt{2}$. In particular, the small-$\xi$ outer solution is $\hat{X} = 1 - \xi/\sqrt{2}$ at leading order. Coming back to $x = (1 - X)/(1 - \varepsilon)$, this asymptotic analysis proves that $x \simeq \xi/\sqrt{2}$ for $\xi < \sqrt{2}$ and $x \simeq 1$ for $\xi > \sqrt{2}$, notwithstanding the transition layer close to $\xi = \sqrt{2}$.

To introduce an effective diffusivity, we need to compute $\zeta = \int_0^\xi [\ell(u)/\ell_0] \mathrm{d}u$. Now, Eq.~[\ref{Eq:l_i}] shows that $\ell/\ell_0 = \varepsilon/X$. Hence, from the results of the asymptotic analysis, $\ell/\ell_0 \simeq \varepsilon (1 - \xi/\sqrt{2})^{-1}$ for $\xi < \sqrt{2}$, and $\ell/\ell_0 \simeq 1$ for $\xi > \sqrt{2}$. Hence, if $\xi < \sqrt{2}$, $\zeta \simeq -\varepsilon\sqrt{2} \, \ln (1 - \xi/\sqrt{2})$. Inverting this relation yields $\xi = (1 - \mathrm{e}^{-\zeta/\varepsilon\sqrt{2}})\sqrt{2}$. Now, in the considered range $\xi < \sqrt{2}$, we have proven that $x \simeq \xi/\sqrt{2}$. This finally yields:
\begin{equation} \label{Eq:x_zeta}
    x = 1 - \mathrm{e}^{-\zeta/\varepsilon\sqrt{2}}.
\end{equation}
This functional form differs from the solution of the linear diffusion problem $x_{\mathrm{hom}}(D,\zeta) = \erf(\zeta/2\sqrt{D})$. We can make an estimate of the effective diffusivity as a best fitting parameter, using the classical least-square method:
$$ \bar{D}_{\mathrm{eff}} = \arg\min_D \int_0^\infty [x_{\mathrm{hom}}(D,\zeta) - x(\zeta)]^2 \mathrm{d}\zeta , $$
where $x(\zeta)$ is given by [\ref{Eq:x_zeta}]. The integral gives an analytical function, but its minimum must be found numerically; it yields:
$$ \bar{D}_{\mathrm{eff}} = 1.40\varepsilon^2 , $$
which is the final result of this asymptotic analysis.

\subsection*{Bubble flux}

We justify Eq. \ref{eq:flux} that provides an expression for  $\Theta^+_{i,k}(t)$,  the fraction of bubbles initially in layer $i$ that have moved at least to layer $k$, by a recursive demonstration starting from the first layers. We have introduced in the main text the number of bubbles $N_i=w \ell_i/\mathcal{A}_i=N_0 \ell_0/\ell_i$ that must be contained in a layer of thickness $\ell_i$. At a given time $t$, the layer $i$ has thus absorbed $N_i(t)-N_0=N_0(\ell_0/\ell_i(t)-1)\equiv N_0 n_i(t)$ additional bubbles. To evaluate the change of topological distance induced by this recruitment of additional bubbles in shrinking layers, we assume a sequential evolution of the bubbles, i.e. bubbles of layer $i$ having reached layer $k$ will only move to layer $k-1$ (if needed) when all bubbles initially in layer $j$, $k-1<j < i$ will already have done so. For bubbles initially in layer 2, their topological distance can be modified because they are absorbed by layer 1 ; however, each bubble can lose one topological distance only once, i.e. the number of bubbles from layer 2 that will complete layer 1 is bounded by the total available number of bubbles in layer 2. In full layer unit, it means that the potential number $n_1$ of bubbles leaving layer 2 is bounded by 1. Assuming an instantaneous motion, we have therefore:
\begin{equation}\label{eq:t21} \Theta^+_{2,1}(t)=\min[1,n_1(t)]. \end{equation} 

Bubbles initially in layer 3 can move to layer 2 either directly because of layer 2 requiring $n_2$ bubbles or indirectly because of bubbles of layer 2 becoming bubbles of layer 1.  This leads to 

\begin{equation}\label{eq:t32} \Theta^+_{3,2}(t)=\min[1,n_1(t)+n_2(t)]. \end{equation}

Additionally, some bubbles having reached layer 2 may go to layer 1 if the requirement of $n_1$ additional bubbles in this layer has not been fulfilled by the bubbles initially in layer 2, i.e., if $n_1>1$. The number  of bubbles from layer 3 that could complete layer 1 is thus $\max[0, n_1(t)-1]$, i.e., it is 0 as long as $n_1(t)<1$, when the layer 1 is fed by bubbles initially from layer 2, and potentially non zero when the bubble requirement of layer 1 has not been fulfilled already. It is also bounded by 1. Hence, 

\begin{equation}\label{eq:t31} \Theta^+_{3,1}(t)=\min\{ 1,\max[0, n_1(t)-1] \}. \end{equation}

Following the same principles, we obtain the general formula \begin{equation}
 \Theta^+_{i,k}(t)   =\min\left\{ 1,\max\left[ 0,\sum_{j=1}^k n_j(t) -(i-1-k)\right] \right\} . \label{eq:flux-method} \end{equation}
This equation is valid for all $i\ge 1$ and $k \le i$ and includes the trivial cases $\Theta^+_{1,k}=0$ and $\Theta^+_{i,i} = 1$.

With this expression established one can then evaluate e.g. the fraction  $\Theta_{i,k}(t)$ of the $N_0$ bubbles of initial layer $i$ that are in layer $k$, as $ \Theta_{i,k}(t) =\Theta^+_{i,k}(t) -\Theta^+_{i,k-1}(t)$, with the obvious convention $\Theta^+_{i,0}=0$, or the average topological distance of the bubbles of initial layer $i$, as 
\begin{equation}\label{eq:Ii-meth} I_{i}(t)=\sum_{k=1}^i k\, \Theta_{i,k}(t)= i-\sum_{k=1}^{i-1} \Theta^+_{i,k}(t).\end{equation} 
\section*{Acknowledgments}
We thank Rub\'en Espeleta Bol\'ivar, J\'er\^ome Giraud, Philippe Moreau and Irène Ventrillard for experimental help. This work was supported by Région Auvergne-Rhône-Alpes (France) through the grant SELFI - Pack Ambition Recherche

\bibliography{biblio_front}
\end{document}